\documentclass{jpsj3}
\usepackage{txfonts}
\usepackage{color}
\usepackage{braket}
\usepackage{bm}

\title{Narrow Zero Mode in Organic Massless Dirac Electron System $\alpha$-(BEDT-TTF)$_2$I$_3$}

\author{Ayano Mori$^1$, Yoshitaka Kawasugi$^1$, Ryusei Doi$^2$, Toshio Naito$^2$, Reizo Kato$^3$, Yutaka Nishio$^1$, and Naoya Tajima$^1$\thanks{naoya.tajima@sci.toho-u.ac.jp}}

\inst{
$^1$Department of Physics, Toho University, Funabashi, Chiba 274-8510, Japan\\ 
$^2$Graduate School of Science and Engineering, Ehime University, Matsuyama 790-8577, Japan \\
$^3$RIKEN, Hirosawa 2-1, Wako-shi, Saitama 351-0198, Japan
}

\abst{
We investigated the interlayer magnetoresistance in an organic massless Dirac electron system $\alpha$-(BEDT-TTF)$_2$I$_3$ under pressure. 
We experimentally demonstrate that the width of the zero mode owing to carrier scattering is much narrower than that of the other Landau levels. 
}

\allowdisplaybreaks

\newcommand{\be}{ \begin{equation}}
\newcommand{\ee}{ \end{equation}}

\usepackage{color}

\usepackage{ulem}

\begin{document}
\maketitle

A rich variety of materials with massless Dirac electrons have been established after the discovery of graphene\cite{rf:1, rf:2}. 

The massless Dirac electron system exhibits a unique energy structure, with two cone-type bands touching each other at discrete points called Dirac points. 
The low-energy excitations described by the relativistic Dirac equation generate the $\pi$ Berry phase that characterizes the transport. 


We can detect the effects of the $\pi$ Berry phase on transport in a magnetic field\cite{rf:1, rf:2}. 
In a magnetic field, the relativistic Landau level (LL) structure is expressed as $E_{N}=\pm \sqrt{2e\hbar v_{\rm F}^2 |N||B|}$, where $v_{\rm F}$ is the Fermi velocity, $N$ is the Landau index, and $B$ is the magnetic field strength. 
The $\pi$ Berry phase yields a phase factor $\pi$ for quantum oscillations originating from the relativistic LL structure. 
The most fascinating phenomenon is the discovery of the half-integer quantum Hall effect (QHE). 
$N=0$ LL, called the zero mode, always exists at Dirac points in a magnetic field and is a remarkable characteristic of the $\pi$ Berry phase.

Moreover, the $\pi$ Berry phase suppresses the backward scattering of electrons. 
Therefore, when the Fermi energy is near Dirac points, the system has extremely high carrier mobility\cite{rf:3}.

Now, our question is, "Does the zero mode exhibit broadening due to carrier scattering?"
In general, each LL in a two-dimensional (2D) electron system under a magnetic field has a finite energy width, $\Gamma_N$, owing to carrier scattering or thermal energy $k_{\rm B}T$. 
When $k_{\rm B}T\gg 2\hbar/\tau$, then $\Gamma_N \sim k_{\rm B}T$, where $\tau$ is the carrier lifetime. At low temperatures where $k_{\rm B}T\ll 2\hbar/\tau$, $\Gamma_N \sim 2\hbar/\tau$. 

In graphene with ripples, it was experimentally demonstrated that the broadening of the zero mode is much narrower than that of other LLs\cite{rf:4}. 
The random magnetic fields caused by ripples broaden the higher LL states; however, the zero-mode state is unaffected.
According to the theory for the case of clean graphene with ripples, the preserved chiral symmetry and suppressed inter-valley scattering do not broaden to the zero mode\cite{rf:5}.

However, the broadening of the zero mode is affected by the random disorder that breaks the chiral symmetry or charge inhomogeneity\cite{rf:6}. 

In this study, we investigate the broadening of the zero mode in a multilayered massless Dirac electron system $\alpha$-(BEDT-TTF)$_2$I$_3$ [BEDT-TTF: bis(ethylenedithio)tetrathiafulvalene] under pressure\cite{rf:7, rf:8, rf:9}. 
This system provides the testing ground for the examination of the broadening of LLs due to carrier scattering because the effects of ripples and charge inhomogeneity in a crystal can be ignored.
Moreover, because the Fermi energy of this system is at Dirac points, the effects of the zero mode on the magnetotransport are detected \cite{rf:10, rf:11, rf:12, rf:13, rf:14, rf:15, rf:16}. 

In this paper, we experimentally demonstrate that the broadening of the zero mode in this system is much narrower than that of the other LLs.

The interlayer resistance, $R_{zz}$, of the $\alpha$-(BEDT-TTF)$_2$I$_3$ crystal under hydrostatic pressures of up to 1.7 GPa, which were applied using a clamp-type pressure cell made of hard alloy MP35N, was measured using a conventional dc method with an electric current of 1 $\mu$ A along the $c$-crystal axis normal to the 2D plane. 

Examination of $R_{zz}$ under a transverse magnetic field (inset of Fig. 1(a)) is one of the most powerful tools for detecting zero-mode effects\cite{rf:10, rf:11, rf:12}. 
In this field configuration, the electric current and magnetic field exhibit weak interactions because they are parallel. 
On the other hand, the magnetic field yields zero-mode carriers. 
The effect of the magnetic field appears only through the change in the carrier density at the vicinity of the Dirac points. 

There are some reports on $R_{zz}$ in this system\cite{rf:10, rf:11, rf:12}. 
This interpretation is based on the assumption that the broadening of the zero mode is the same as that of the $N \neq 0$ LLs. 
However, we found that the width of the zero mode is much narrower than that of the other LLs from a detailed reinterpretation, as follows:

\begin{figure}
\begin{centering}
\includegraphics[trim=0 0 0 0, width=7.8cm, angle=0, clip]{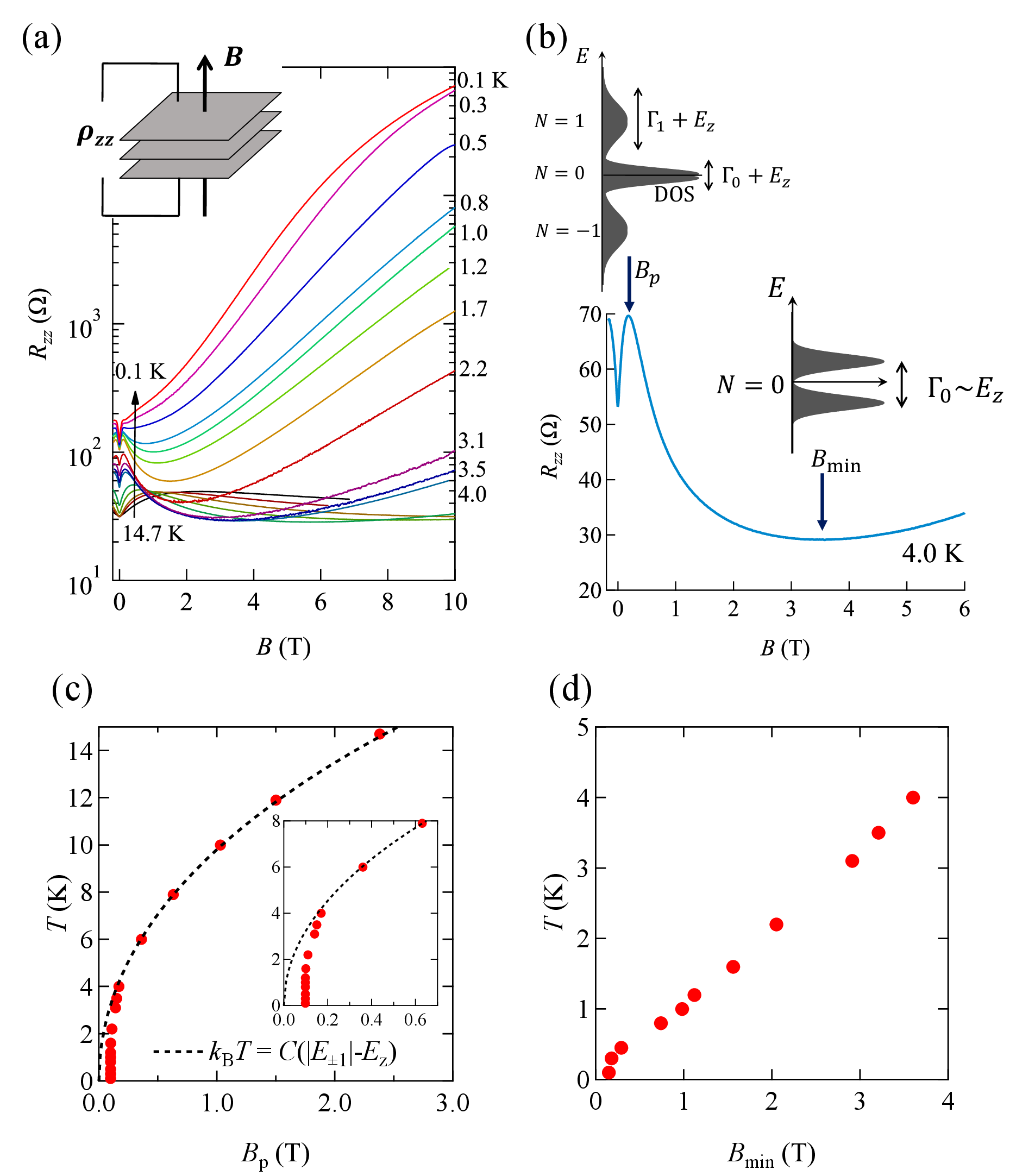}
\caption{\label{fig1}(Color online) 
Magnetic field dependence of $R_{zz}$ at temperatures below 14.7 K. The inset shows the schematic of the experiment for $R_{zz}$ under a transverse magnetic field. (b) Magnetic field dependence of $R_{zz}$ at 4.0 K. The insets show schematics of the zero mode and $N=\pm1$ LL. (c) Temperature plot against $B_{\rm p}$. The curve of Eq. (1) is indicated by the dashed line. The inset shows the region at low temperature and low field. (d) Temperature plot against $B_{\rm min}$.
}
\end{centering}
\end{figure}

The magnetic field dependence of $R_{zz}$ at temperatures below 14.7 K is shown in  Fig. 1(a). 
The results reproduce our previous data published in Ref. 11. 
We can understand $R_{zz}$ well in terms of zero-mode carriers, including their Zeeman splitting. 
For example, $R_{zz}$ at 4.0 K is enlarged in Fig. 1(b). 
The magnetic field effects on the zero mode and other LLs result in the formation of a peak and a minimum of $R_{zz}$, as indicated by the arrows in Fig. 1(b).

In the following, we examine the zero mode and other LL broadening from the interpretation of the peak and minimum of $R_{zz}$ in Fig. 1.

First, we focus on the $R_{zz}$ peak.
Because each LL is broadened by the scattering of carriers and/or thermal energy,  the zero mode overlaps with other LLs at low magnetic fields. 
The overlap between the zero mode and other LLs, primarily $N=\pm1$ LL, is sufficiently small above the critical field $B_{\rm p}$. Most of the mobile carriers are in the zero mode, and thus, the negative magnetoresistance is dependent on the degree of degeneracy observed, as shown in Fig. 1.
Such a situation is called the quantum limit. 
When $k_{\rm B}T\gg 2\hbar/\tau$, we have a tentative relationship,
\begin{equation}
k_{\rm B}T=C(\sqrt{2e\hbar v_{\rm F}^2 B_{\rm p}}-E_z),
\label{eqn:1}
\end{equation}
where $C$ denotes the scale factor, $E_z=2\mu_{\rm B} B_{\rm p}$ denotes the Zeeman energy, and $\mu_{\rm B}$ denotes the Bohr magneton\cite{rf:12, rf:16, rf:17}. 
In fact, Eq. (1) with $C=0.6$ and $v_{\rm F} = 4.5\times 10^4$ m/s\cite{rf:18} is reproduced well, except in the temperature region below 4 K, as shown in Fig. 1(c). 
Note that the above is the same interpretation as in previous studies \cite{rf:12, rf:16, rf:17}. The following is a new analysis:

In the temperature region below 4 K, the broadening of LLs is due to the scattering of carriers so that $B_{\rm p}$ becomes constant at approximately 0.12 T, as shown in the inset of Fig. 1(c).
Thus, Eq. (1) is rewritten as
\begin{equation}
\displaystyle \Gamma_0/2+\Gamma_{\pm 1}/2+E_z=C(\sqrt{2e\hbar v_{\rm F}^2 B_{\rm p}}-E_z).
\label{eqn:2}
\end{equation}
Then, we have
\begin{equation}
(\Gamma_0+\Gamma_{\pm 1})/2 \sim 3.9k_{\rm B}. 
\label{eqn:3}
\end{equation}
This energy is consistent with the temperature at which Eq. (1) lacks validity, as shown in Fig. 1(c).

Assuming that $\Gamma_0=\Gamma_{\pm 1}$, we obtain $\Gamma_0/k_{\rm B}=\Gamma_{\pm1}/k_{\rm B}=3.9$ K at low temperatures. 
However, in the next step, we demonstrate that $\Gamma_0$ is much narrower than this value.

To examine $\Gamma_0$, we examine the case of $R_{zz}$ minimum.
In the quantum limit state, the degeneracy of the zero mode is increased in proportion to the field, and a negative magnetoresistance is observed, as shown in Fig. 1(a) and (b).
The Zeeman effect, on the other hand, has a significant influence on transport at a high magnetic field.
Each LL releases the degeneracy and splits it into two levels with energies $E_{N}\pm E_z/2$.
This effect on the zero-mode carrier density is the strongest because the levels are shifted from the Fermi energy (Dirac points). 
At a high field, where $\Gamma_0 < E_z$, reducing the density of zero-mode carriers increases $R_{zz}$. 
At $R_{zz}$ minimum, $\Gamma_0 \sim E_z = 2\mu_{\rm B}B_{\rm min}$\cite{rf:10}. 

A plot of temperature against the magnetic field $B_{\rm min}$ at the $R_{zz}$ minimum is shown in Fig. 1(d). 
The linearity between the temperature and $B_{\rm min}$ down to 0.3 K indicates $\Gamma_0 \sim k_{\rm B}T$.
At temperatures lower than 0.3 K, the data deviate from linearity because the zero mode broadens mainly due to carrier scattering.
Thus, the broadening of the zero mode due to carrier scattering is estimated to be $\Gamma_0/k_{\rm B} \sim 0.2$ K from the relationship $\Gamma_0 \sim 2\mu_{\rm B}B_{\rm min}$ at 0.1 K. 
Note that this value is less than half of that reported in Ref. 11. The difference in the amount of doping due to the unstable I$_3$$^-$ anion\cite{rf:9} may be responsible for this discrepancy.

Lastly, combining Eq. (3) with $\Gamma_0/k_{\rm B}T \sim 0.2$ K, we obtain $\Gamma_{\pm 1}/k_{\rm B} \sim 8$ K at 0.1 K. We compared this value with that of the zero-mode.
Note that the broadening of the $N=\pm 1$ LL is consistent with the Dingle temperature in quasi-2D organic metals\cite{rf:19}.
In this sense, the zero mode is a special LL. 
The suppression of the inter-valley scattering does not give the broadening.

We obtain a long lifetime $\tau \sim 7 \times 10^{-11}$ s and a long mean free path $l=v_{\rm F} \tau \sim 3$ $\mu$m from the relationship $\Gamma_0 = 2\hbar/\tau$ at 0.1 K.

In conclusion, we examined the broadening of the zero mode and $N=\pm1$ LL by investigating the interlayer magnetoresistance in an organic massless Dirac electron system $\alpha$-(BEDT-TTF)$_2$I$_3$ under pressure. 
The broadening of the zero mode at low temperatures is extremely narrow, $\Gamma_0 \sim 0.2$ K at 0.1 K. 
On the other hand, the $N=\pm 1$ LLs have much wider broadening, $\Gamma_{\pm 1} \sim 8$ K.

\acknowledgments{This work was supported by MEXT/JSPJ KAKENHI under Grant No. 20K03870.}


\end{document}